\documentclass[aps,prl,twocolumn,groupedaddress]{revtex4}
\usepackage{amsmath}
\usepackage{amssymb}
\usepackage{mathbbol}
\usepackage{graphicx}
\usepackage{psfrag}
\usepackage{color}
\usepackage[colorlinks=true,linkcolor=red,citecolor=blue,urlcolor=green]{hyperref}

\begin{document}

\hyphenation{Ryd-berg}

\newcommand{\fig}{\text{Fig.~}}

\newcommand{\lsim}{\mathrel{\hbox{\rlap{\lower.55ex\hbox{$\sim$}} \kern-.3em \raise.4ex \hbox{$<$}}}}

\title{An echo experiment in a strongly interacting Rydberg gas}

\author{Ulrich Raitzsch}
\email[Electronic address:]{u.raitzsch@physik.uni-stuttgart.de}
\author{Vera Bendkowsky}
\author{Rolf Heidemann}
\author{Bj\"{o}rn Butscher}
\author{Robert L\"{o}w}
\affiliation{5. Physikalisches Institut, Universit\"{a}t
Stuttgart, Pfaffenwaldring 57, 70569 Stuttgart, Germany}
\author{Tilman Pfau}
\email[Electronic address:]{t.pfau@physik.uni-stuttgart.de}
\affiliation{5. Physikalisches Institut, Universit\"{a}t
Stuttgart, Pfaffenwaldring 57, 70569 Stuttgart, Germany}

\date{\today}

\begin{abstract}
When ground state atoms are excited to a Rydberg state, van der Waals interactions among them can lead to a strong suppression of the excitation. Despite the strong interactions the evolution can still be reversed by a simple phase shift in the excitation laser field. We experimentally prove the coherence of the excitation in the strong blockade regime by applying an `optical rotary echo' technique to a sample of magnetically trapped ultracold atoms, analogous to a method known from nuclear magnetic resonance. We additionally measured the dephasing time due to the interaction between the Rydberg atoms.
\end{abstract}

\maketitle

Interactions are often characterized by a length scale $a$. In the case of strong interaction $a$ is much larger than the mean distance $r$ between two particles ($r \sim n^{-1/3}$, where $n$ is the particle density). Contrary to dilute systems, where $na^3 \ll 1$, strongly interacting systems are dominated by correlations that make mean field descriptions inaccurate or even misleading.

Strongly correlated \emph{ground states} of ultracold fermionic quantum systems are currently studied in the BEC-BCS crossover regime \cite{Gri06}. Here, the length scale for the interaction, the s-wave scattering length $a$ is increased by means of a Feshbach resonance. Superfluid behavior has been observed even in the strongly interacting regime \cite{Zwi05}. Likewise, the Mott insulator state is a strongly correlated state of matter which is studied using ultracold gases in optical lattices \cite{Gre02}.

\emph{Excited states} can also  be correlated by strong interactions, leading to collective states and blockade phenomena. Several proposals have been made on how to use this blockade mechanism for the implementation of fast quantum gates for thermally frozen neutral atoms \cite{Jak00, Luk01}. In the case of the excitation of gaseous atoms to a Rydberg state by a laser with a small linewidth, a length scale for the interaction among them can be found by equating the mutual interaction energy with the power broadened transition linewidth $\hbar \Omega$, where $\Omega$ is the Rabi frequency. For the purpose of this paper it can be assumed that the Rydberg--Rydberg interaction is dominated by the van der Waals interaction $V(r) =-C_6/r^6$. 
Strong blockade effects are expected, if the resulting blockade radius $a_{\text{block}}\propto\sqrt[6]{|C_6|/\hbar \Omega}$ is much larger than the mean distance of the ground state atoms, given by the density of ground state atoms $n_\text{g}^{-1/3}$. Such blockade effects have been studied for the van der Waals interaction  and for resonant dipolar interactions between Rydberg atoms \cite{Sin04, Ton04, Cub05, Vog06}. These experiments were done in a regime where $n_{\text{g}} a_{\text{block}}^3$ is on the order of 10 or less and the blockade effect can be understood in a mean field model, where the interactions tune the Rydberg state  out of the laser resonance \cite{Ton04,Rob05}.

In the strong blockade regime, where $n_{\text{g}} a_{\text{block}}^3 \gg 1$, the system can be described as two-level atoms and beyond mean field effects have been predicted for samples smaller than $a_{\text{block}}$ \cite{Luk01,Rob05}. The excitation is then limited to a single excitation into a Rydberg state, delocalized over all $N$ members of a blockade sphere with radius $a_{\text{block}}$. The corresponding collective transition matrix element between the ground state and the state carrying a single excitation is given by  $\sqrt{N} \Omega_0$ \cite{Luk01}, where $\Omega_0$ is the single atom Rabi frequency. These objects are sometimes called `superatoms'  \cite{Vul06} as the $N$ members of the ensemble resemble a two-level system with the enhanced collective Rabi frequency $\sqrt{N}\Omega_0$.

In previous experiments conducted in the strong blockade regime, we have confirmed the collective coherent scaling of the excitation rates with $\sqrt{N}\Omega_0$ \cite{Hei07}. This letter aims at a direct demonstration of the coherence of the excitation and the measurement of the dephasing times in a strongly interacting Rydberg gas, where $n_{\text{g}} a_{\text{block}}^3$ is a few thousand. 

For a thermal atomic cloud, trapped in a harmonic potential, the density $n_{\text{g}}$ has a Gaussian distribution over the position in the sample. Thus, the atom number $N$ and, consequently, the collective Rabi frequency $\sqrt{N}\Omega_0$ is inhomogenously distributed. This leads to the fact that Rabi oscillations cannot be observed directly, because an integrated signal over all Rabi frequencies is detected. Echo techniques, such as the rotary echo sequence  \cite{Sol59}, have been developed in nuclear magnetic resonance physics to overcome these inhomogeneity effects. 

In frozen Rydberg gases the center of mass motion of the atoms can be neglected. If the narrowband excitation laser is tuned to resonance, the Hamilton operator can be written using the Pauli matrices $\sigma_x$ and $\sigma_z$ \cite{Rob05} as:

\begin{gather}
    \label{HamOp}
 \mathcal{H} = \sum\limits_j \Omega_0 \left(\sigma_x\right)_j + \sum\limits_{j<k} V_{jk}\,\frac{1}{2}\left(\mathbb{1} + \sigma_z\right)_j\,\frac{1}{2}\left(\mathbb{1} + \sigma_z\right)_k ,
\end{gather}
where $V_{jk}$ is the pair interaction between two Rydberg atoms $j$ and $k$. This model assumes for single atoms a two-level system consisting of a ground and a Rydberg state at site $j$.

In a rotary echo \cite{Sol59} sequence an atom is excited and after a certain time $\tau_{\text{p}}$ the sign of the excitation amplitude $\Omega$ is reversed. Independent of the magnitude of $\Omega$, the state of the system evolves back to the electronic ground state at time $2 \tau_{\text{p}}$, unless dephasing occurs, e.g., due to the interaction $V_{jk}$.

Strong interactions $V_{jk}$ also strongly suppress the excitation and  build up spatial correlations in the distribution of Rydberg states \cite{Rob05}. Therefore a mean field model is not expected to lead to a correct description of the dephasing. Spatial correlations are taken into account in a model that is based on the collective superatom states. Such an approach is more appropriate although of course not complete. A superatom model would predict a coherent evolution with reduced dephasing compared to a mean field model as the spatial correlations prevent short distances between Rydberg atoms. Therefore, even in the case of strong suppression of excitation an echo signal is expected.

\begin{figure}[t]
            \includegraphics[width=86mm]{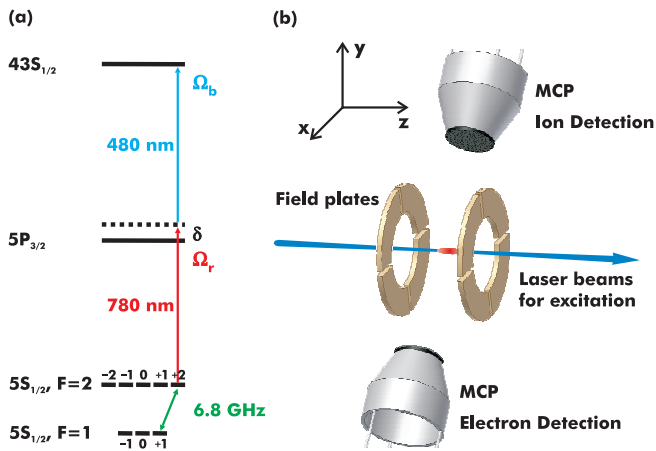}
        \caption{\label{Fig:Setup}(a) Level scheme for the two-photon transition and the Landau-Zener sweep. For the purpose of the Landau-Zener sweep the two hyperfine levels in $^{87}$Rb are coupled. (b) Setup used for the experiments. The excitation laser represents the two superimposed laser beams for the two-photon transition. The atomic cloud (red) sits in the middle between eight field plates, which can independently be set to high voltages. Two multi channel plates (MCP) are used for detection of ions and electrons, respectively.}
\end{figure}

To measure the dephasing in the excitation of a strongly interacting Rydberg gas, a sample of Rubidium atoms in a magnetic trap is prepared (\fig{\ref{Fig:Setup}} and \cite{Loe07}). The atoms are cooled to a temperature of 3.8 $\mu$K at a peak density of ground state atoms $n_{\text{g}}$ of $5.2\times 10^{19}\text{ m}^{-3}$. For a systematic variation of the density, a Landau-Zener sweep is used \cite{Rub81}. A microwave sweep of 6.8 GHz couples the two hyperfine ground states of $^{87}$Rb and, depending on the sweep time, a fraction of the atoms are transfered from the initial 5S$_{1/2}, F=2, m_F = 2$ state into the untrapped $F=1, m_F = 1$ state (\fig{\ref{Fig:Setup}}(a)). The size of the sample depends only on the temperature and the trap frequencies, which are held constant for all data presented here. The sample therefore has a constant size and is fully characterized by either the number of ground state atoms $N_{\text{g}}$ or the peak density $n_{\text{g}}$ and $T$. The number of ground state atoms $N_{\text{g}}$ in the trap was varied between $1.1\times 10^7$ and $1.0\times 10^6$ atoms without a significant change in temperature $T$.

Afterwards, the atoms are excited by a resonant two-photon transition into the 43S$_{1/2}$ Rydberg state, using 480 nm and 780 nm light, which is detuned by 2$\pi\times$472 MHz from the 5P$_{3/2}$ intermediate state (\fig{\ref{Fig:Setup}}(a) and \cite{Loe07}). A value for the two-photon Rabi frequency $\Omega_0$ of 2$\pi\times$90.5 kHz is found using a calculation of the dipole matrix element between the 5P$_{3/2}$ and the 43S$_{1/2}$ state and a value for $\Omega_{\text{r}}$ (\fig\ref{Fig:Setup}(a)) of 8.7 MHz. The obtained value is in excellent agreement with values from independent calculations \cite{Rob07pc}. The Rabi frequency can be treated as constant over the atomic cloud, as the cloud is smaller than the waists of the exciting lasers. The two laser beams are superimposed along the $z$-direction of the magnetic trap, i.e., the long axis of the cigar shaped atomic cloud (\fig\ref{Fig:Setup}(b)). The magnetic offset field in $z$-direction defines the quantization axis, so that, by choosing $\sigma^+$ for the 780 nm and $\sigma^-$ light for the 480 nm laser, the atoms are selectively excited into the 43S$_{1/2},\,m_J = +1/2$ state, which has the same Zeeman energy shift as the ground state. In order to exclude the effects of ions, an electric field of 200 V/m is applied during the excitation \cite{Hei07}. Subsequent to the excitation, the Rydberg atoms are field ionized and pushed towards a micro channel plate by applying an electric field of 20 kV/m. Temperature and density of the atomic ensemble are obtained from absorption images of the remaining ground state atoms.

\begin{figure}[ht]
\includegraphics[width=86mm]{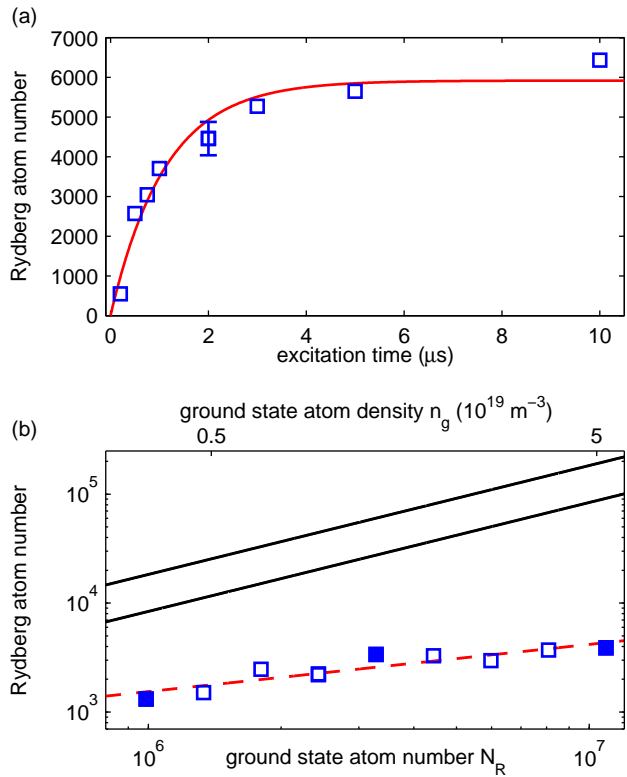}
 \caption{\label{Fig:Int_vs_NoInt}(a) Typical Rydberg atom numbers as a function of excitation time $\tau$ for $n_{\text{g}}=5.2\times 10^{19}\text{ m}^{-3}$, i.e., $N_{\text{g}}=1.1\times 10^7$ and $\Omega_0 = 2\pi\times$90.5 kHz. The error bar shows the statistical fluctuation of the Rydberg atom number over ten experiments at every excitation time. (b) Rydberg atom number $N_{\text{R}}$ (\textcolor{blue}{$\square$}) as a function of the number of ground state atoms $N_{\text{g}}$ and their density $n_{\text{g}}$ (upper scale). From a power-law fit $N_{\text{R}} \propto N_{\text{g}}^{0.43\pm 0.03}$ is deduced. For comparison the upper solid line shows the expected Rydberg number for an excitation of non-interacting Rydberg atoms. The lower solid line takes a frequency uncertainty of the excitation laser of 2$\pi\times$1.5 MHz into account. The filled data points (\textcolor{blue}{$\blacksquare$}) indicate the atom numbers for which the rotary echo signals are shown in \fig\ref{Fig:EchoExp}. The statistical fluctuation is smaller than the marker size.}
\end{figure}

\begin{figure}[t]
\includegraphics[width=86mm]{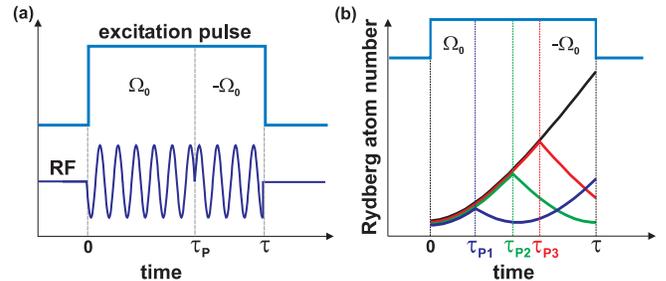}
 \caption{\label{Fig:EchoSchematic}
 Schematic of the pulse sequence for the rotary echo experiment. (a) An excitation pulse of constant duration $\tau$ is applied. After a variable time $\tau_{\text{p}}$ the phase of the radio frequency of the AOM, which switches the 480 nm light, is shifted by $\pi$. The phase shift is equivalent to inverting the sign of $\Omega_0$ to $-\Omega_0$. (b) The coloured curves schematically show the excitation dynamics for the whole sample during the pulse sequence for some values of $\tau_p$ if no dephasing due to interaction occurred. The black line indicates the time evolution of the Rydberg population without changing the sign of the Rabi frequency. This corresponds to the evolution of excitation in \fig\ref{Fig:Int_vs_NoInt}(a) for short excitation times, whereas the initial quadratic rise is exaggerated and typically changes to a linear rise after a few ten nanoseconds.}
\end{figure}

\begin{figure}[ht]
\includegraphics[width=86mm]{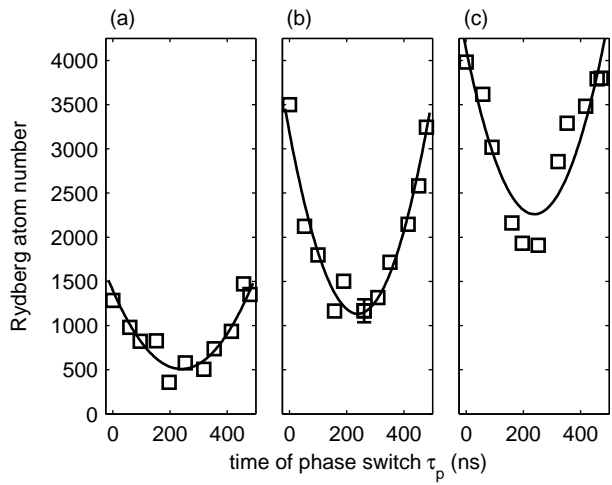}
 \caption{\label{Fig:EchoExp}
 Three typical rotary echo measurements for different ground state atom densities. During an excitation pulse of constant duration $\tau$ a $\pi$-phase flip of the radio frequency of the AOM, which controls the 480 nm light, is applied at a variable time $\tau_{\text{p}}$. This results in a change of the sign of $\Omega_0$ to $-\Omega_0$. By this the excitation is, in the absence of any dephasing effect, completely reversed. A parabolic function is fitted to the data to guide the eye. The visibilities of the rotary echo signal obtained are $(47\pm8)$ \% with $N_{\text{g}} = 1.0\times 10^{6}$ in (a), $(48\pm5)$ \% with $N_{\text{g}} = 3.3\times 10^{6}$ in (b) and $(29\pm6)$ \% with $N_{\text{g}} = 1.0\times 10^{7}$ in (c). The error bar indicates the statistical fluctuation of the Rydberg atom number over five independent experiments.}
\end{figure}

A typical excitation curve is shown in \fig\ref{Fig:Int_vs_NoInt}(a). Such excitation curves were systematically studied in our previous work \cite{Hei07}. For the purpose of this work, a constant pulse with a duration $\tau$ is applied and the number of Rydberg atoms for different numbers of ground state atoms in the sample is investigated. For the measurements in \fig{\ref{Fig:Int_vs_NoInt}(b), \ref{Fig:EchoSchematic}, \ref{Fig:EchoExp}} the time $\tau$ was held constant at 478 ns. The pulse duration $\tau$ was chosen to be short compared to the time scale of the saturation, which is on the order of microseconds (\fig\ref{Fig:Int_vs_NoInt}(a)). Figure \ref{Fig:Int_vs_NoInt}(b) shows that the experiments are performed in the strong interaction regime: Without interaction one would expect Rydberg atom numbers $N_{\text{R}}$ depicted by the two solid lines. The upper bound is given by $N_{\text{R}} = N_{\text{g}}\, \sin^2(\Omega_0\,\tau/2) = 0.02\, N_{\text{g}}$. Taking a statistically distributed laser detuning with an experimentally determined width of $2\pi\times$1.5 MHz into account, the number of Rydberg atoms would decrease to the value $N_{\text{R}} = 0.01\, N_{\text{g}}$. In both cases the Rydberg atom number scales linearly with the number of ground state atoms $N_{\text{g}}$. In comparison, our data show significantly less excitation and an increase of $N_{\text{R}}$, which scales only like $N_{\text{g}}^{0.43 \pm 0.03}$. Thus, a strong suppression of the Rydberg atom number due to the van der Waals interaction is observed. This clearly indicates that the following experiments are done in the strong blockade regime, where strong interactions among the Rydberg atoms suppress the excitation substantially.

In order to demonstrate coherence in the sense of a unitary evolution of the system, the experimental sequence is changed. After preparing the ground state atoms in the magnetic trap, a square laser pulse with a pulse duration $\tau$ and a Rabi frequency $\Omega_0$ is switched on. After the time $\tau_{\text{p}} \leq \tau$ the phase of the radio frequency of the AOM, which is used to control the 480 nm light, is shifted by $\pi$ (\fig\ref{Fig:EchoSchematic}(a)). This phase flip changes the Rabi frequency from $\Omega_0$ to $-\Omega_0$.
With this rotary echo sequence, the ground state atoms are excited for a time $\tau_{\text{p}}$ and de-excited for a time $\tau-\tau_{\text{p}}$ if no substantial dephasing, e.g. by an inhomogeneous field, occurred (\fig\ref{Fig:EchoSchematic}(b)).

Three typical rotary echo measurements are shown in \fig\ref{Fig:EchoExp} for a number of ground state atoms $N_{\text{g}} = 1.0\times 10^{6}$ in (a), $N_{\text{g}} = 3.3\times 10^{6}$ in (b) and  $N_{\text{g}} = 1.0\times 10^{7}$ in (c). For every rotary echo curve, five independent atomic ensembles have been prepared and the time $\tau_{\text{p}}$ has been scanned from 0 to $\tau$ in 10 steps within every sample. The excitation was resonant with the two-photon transition for all measurements. The laser frequency has a Gaussian distribution with a width of 2$\pi\times$130 kHz on the time scale on which $\tau_{\text{p}}$ is scanned and a width of 2$\pi\times$1.5 MHz on the long time scale, i.e., between the preparation of two samples. The data in \fig\ref{Fig:EchoExp} has been fitted with a parabolic function to guide the eye and to obtain the visibility

\begin{gather}
 V = \frac{N_{\text{R}}(\tau_{\text{p}}=0)-N_{\text{R}}(\tau_{\text{p}}=\tau/2)}{N_{\text{R}}(\tau_{\text{p}}=0)+N_{\text{R}}(\tau_{\text{p}}=\tau/2)}
\end{gather}

of the rotary echo signal. The visibilities are $(47\pm8)$ \%, $(48\pm5)$ \% and $(29\pm6)$ \% in \fig\ref{Fig:EchoExp}(a)-(c), respectively. This echo signal is a clear for the coherence of the excitation, despite strong interactions among the excited atoms.

\begin{figure}[th]
\includegraphics[width=86mm]{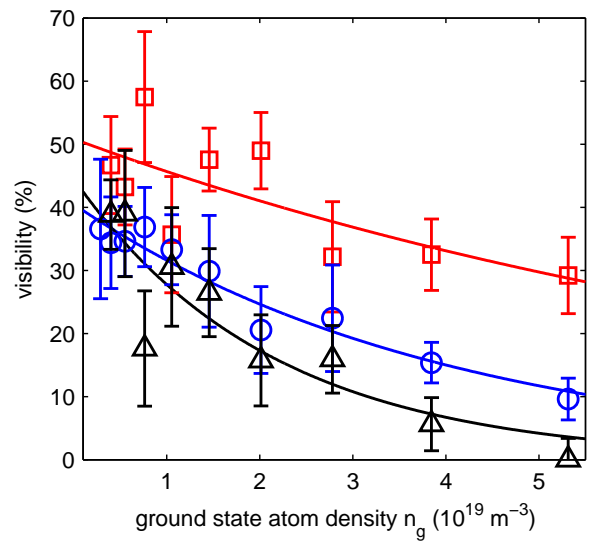}
 \caption{\label{Fig:VisAll}
 Visibility as a function of the peak density of ground state atoms $n_{\text{g}}$ for three different excitation times. Namely, $\tau=478$ ns (\textcolor{red}{$\square$}), $\tau=534$ ns (\textcolor{blue}{$\bigcirc$}) and $\tau=659$ ns ($\triangle$). All three data sets were fitted with a simple exponential decay to guide the eye. The error bars are obtained from the parabolic fits to the data (\fig{\ref{Fig:EchoExp}}).}
\end{figure}

In \fig\ref{Fig:VisAll} we show the dependence of the visibility on the density of ground state atoms for three different pulse lengths. The data was fitted with a simple exponential decay to guide the eye. From the data the tendency of the visibility on the interaction strength among the sample can be obtained:  The higher the density and the longer the pulse length, the smaller the visibility. Assuming an exponential decay for the dependence of the visibility on the pulse time, a dephasing time of 860 ns is obtained for a density $n_{\text{g}} = 7.2\times 10^{18}\text{ m}^{-3}$. Note that the effect due to the radiative lifetime of the Rydberg state is on the order of  100 $\mu$s and, thus, can be safely neglected.

To conclude, we have prepared a cold sample of ground state atoms, from which a few thousand atoms are excited into a strongly interacting Rydberg state. It was shown that even for pulse lengths, short compared to the time scale on which the interaction leads to a full blockade of the Rydberg signal, the excitation was strongly suppressed in comparison to the non-interacting case. The excitation into the Rydberg state was shown to be phase stable for the whole  inhomogeneous sample with respect to the exciting laser field. The phase stability was proven by a so called optical rotary echo technique, where the excitation laser field was phase shifted by $\pi$ during the pulse leading to a coherent de-excitation. The rotary echo signal shows dephasing times of several 100 ns, while $n_{\text{g}}a_{\text{block}}^3 > 1000$. These long dephasing times, despite the strong interactions, give evidence for spatial correlations in the sample. A quantitative comparison with theory requires many-body calculations and will be the subject of further studies.

We acknowledge fruitful discussion with L. Santos, F. Robicheaux and G. Denninger. Financial support is acknowledged from the Deutsche Forschungsgemeinschaft within the SFB/TRR21 and under the contract PF 381/4-1, U.R. acknowledges support from the Landesgraduiertenf\"{o}rderung Baden-W\"{u}rttemberg.\\
Correspondence and request for materials should be addressed to U.R. or T.P.


\begin{thebibliography}{10}
\providecommand{\bibnamefont}[1]{#1}
\providecommand{\bibfnamefont}[1]{#1}
\providecommand{\bibinfo}[2]{#2}

\bibitem{Gri06}
\bibinfo{author}{\bibfnamefont{R.}~\bibnamefont{Grimm}}, in
  \emph{\bibinfo{booktitle}{Ultracold Fermi Gases}}, edited by
  \bibinfo{editor}{\bibfnamefont{M.}~\bibnamefont{Inguscio}},
  \bibinfo{editor}{\bibfnamefont{W.}~\bibnamefont{Ketterle}}, \bibnamefont{and}
  \bibinfo{editor}{\bibfnamefont{C.}~\bibnamefont{Salomon}}
  (\bibinfo{address}{Varenna}, \bibinfo{year}{2006}), CLXIV.

\bibitem{Zwi05}
\bibinfo{author}{\bibfnamefont{M.~W.} \bibnamefont{{Zwierlein}}},
  \emph{et~al.}, \bibinfo{journal}{\nat} \textbf{\bibinfo{volume}{435}},
  \bibinfo{pages}{1047} (\bibinfo{year}{2005}).

\bibitem{Gre02}
\bibinfo{author}{\bibfnamefont{M.}~\bibnamefont{Greiner}}, \emph{et~al.},
  \bibinfo{journal}{Nature} \textbf{\bibinfo{volume}{415}}, \bibinfo{pages}{39}
  (\bibinfo{year}{2002}).

\bibitem{Jak00}
\bibinfo{author}{\bibfnamefont{D.}~\bibnamefont{Jaksch}}, \emph{et~al.},
  \bibinfo{journal}{Phys. Rev. Lett.}
  \textbf{\bibinfo{volume}{85}}(\bibinfo{number}{10}), \bibinfo{pages}{2208}
  (\bibinfo{year}{2000}).

\bibitem{Luk01}
\bibinfo{author}{\bibfnamefont{M.~D.} \bibnamefont{{Lukin}}}, \emph{et~al.},
  \bibinfo{journal}{\prl} \textbf{\bibinfo{volume}{87}}(\bibinfo{number}{3}),
  \bibinfo{pages}{037901} (\bibinfo{year}{2001}).

\bibitem{Sin04}
\bibinfo{author}{\bibfnamefont{K.}~\bibnamefont{Singer}}, \emph{et~al.},
  \bibinfo{journal}{\prl} \textbf{\bibinfo{volume}{93}}(\bibinfo{number}{16}),
  \bibinfo{pages}{163001} (\bibinfo{year}{2004}).

\bibitem{Ton04}
\bibinfo{author}{\bibfnamefont{D.}~\bibnamefont{Tong}}, \emph{et~al.},
  \bibinfo{journal}{\prl} \textbf{\bibinfo{volume}{93}}(\bibinfo{number}{6}),
  \bibinfo{pages}{063001} (\bibinfo{year}{2004}).

\bibitem{Cub05}
\bibinfo{author}{\bibfnamefont{T.}~\bibnamefont{{Cubel Liebisch}}},
  \bibinfo{author}{\bibfnamefont{A.}~\bibnamefont{{Reinhard}}},
  \bibinfo{author}{\bibfnamefont{P.~R.} \bibnamefont{{Berman}}},
  \bibnamefont{and}
  \bibinfo{author}{\bibfnamefont{G.}~\bibnamefont{{Raithel}}},
  \bibinfo{journal}{\prl} \textbf{\bibinfo{volume}{95}}(\bibinfo{number}{25}),
  \bibinfo{pages}{253002} (\bibinfo{year}{2005}).

\bibitem{Vog06}
\bibinfo{author}{\bibfnamefont{T.}~\bibnamefont{{Vogt}}}, \emph{et~al.},
  \bibinfo{journal}{\prl} \textbf{\bibinfo{volume}{97}}(\bibinfo{number}{8}),
  \bibinfo{pages}{083003} (\bibinfo{year}{2006}).

\bibitem{Rob05}
\bibinfo{author}{\bibfnamefont{F.}~\bibnamefont{{Robicheaux}}}
  \bibnamefont{and} \bibinfo{author}{\bibfnamefont{J.~V.}
  \bibnamefont{{Hern{\'a}ndez}}}, \bibinfo{journal}{\pra}
  \textbf{\bibinfo{volume}{72}}(\bibinfo{number}{6}), \bibinfo{pages}{063403}
  (\bibinfo{year}{2005}).

\bibitem{Vul06}
\bibinfo{author}{\bibfnamefont{V.}~\bibnamefont{{Vuletic}}},
  \bibinfo{journal}{Nature Physics} \textbf{\bibinfo{volume}{2}},
  \bibinfo{pages}{801} (\bibinfo{year}{2006}).

\bibitem{Hei07}
\bibinfo{author}{\bibfnamefont{R.}~\bibnamefont{Heidemann}}, \emph{et~al.},
  \bibinfo{journal}{\prl} \textbf{\bibinfo{volume}{99}}(\bibinfo{number}{16}),
  \bibinfo{eid}{163601} (\bibinfo{year}{2007}).

\bibitem{Sol59}
\bibinfo{author}{\bibfnamefont{I.}~\bibnamefont{{Solomon}}},
  \bibinfo{journal}{\prl} \textbf{\bibinfo{volume}{2}}, \bibinfo{pages}{301}
  (\bibinfo{year}{1959}).

\bibitem{Loe07}
\bibinfo{author}{\bibfnamefont{R.}~\bibnamefont{L\"{o}w}}, \emph{et~al.},
  \bibinfo{journal}{arXiv:abs/0706.2639}  (\bibinfo{year}{2007}).

\bibitem{Rub81}
\bibinfo{author}{\bibfnamefont{J.~R.} \bibnamefont{{Rubbmark}}},
  \bibinfo{author}{\bibfnamefont{M.~M.} \bibnamefont{{Kash}}},
  \bibinfo{author}{\bibfnamefont{M.~G.} \bibnamefont{{Littman}}},
  \bibnamefont{and}
  \bibinfo{author}{\bibfnamefont{D.}~\bibnamefont{{Kleppner}}},
  \bibinfo{journal}{\pra} \textbf{\bibinfo{volume}{23}}, \bibinfo{pages}{3107}
  (\bibinfo{year}{1981}).

\bibitem{Rob07pc}
\bibinfo{title}{\text{F. Robicheaux private communication}}.

\end{thebibliography}

\end{document}